\documentclass[]{pasj01}

\begin{document}
\Received{}
\Accepted{}
\Published{yyyy/mm/dd}

\newcommand{\red}{\textcolor{red}}
\newcommand{\blue}{\textcolor{blue}}

\title{An enigmatic hump around 30 keV in Suzaku spectra of Aquila X-1 in the Hard State}

\author{Megu \textsc{Kubota}\altaffilmark{1} \altaffilmark{2}}
\altaffiltext{1}{Department of Physics, Tokyo University of Science, 1-3 Kagurazaka, Shinjyuku-ku, Tokyo 162-0825, Japan}
\email{megu@crab.riken.jp}

\author{Toru \textsc{Tamagawa}\altaffilmark{1} \altaffilmark{2}}
\altaffiltext{2}{High Energy Astrophysics Laboratory, RIKEN Nishina Center, 2-1- Hirosawa, Wako, Saitama 351-0198, Japan}

\author{Kazuo \textsc{Makishima}\altaffilmark{3} \altaffilmark{4} \altaffilmark{5}}
\altaffiltext{3}{MAXI Team, RIKEN, 2-1- Hirosawa, Wako, Saitama 351-0198, Japan}
\altaffiltext{4}{Department of Physics, The University of Tokyo, 7-3-1 Hongo, Bunkyo-ku, Tokyo 113-0033, Japan}
\altaffiltext{5}{Kavl: IPMU, The University of Tokyo, 5-1-1 Kashiwa-no-ha, Kashiwa, Chiba 277-8535}
\author{Toshio \textsc{Nakano}\altaffilmark{2}}
\author{Wataru \textsc{Iwakiri}\altaffilmark{3} \altaffilmark{6}}
\altaffiltext{6}{Department of Physics, Faculty of Science and Engineering, Chuo University, 1-13-27 Kasuga, Bunkyo-ku, Tokyo 112-8551, Japan}
\author{Mutsumi \textsc{Sugizaki}\altaffilmark{3}}
\author{Ko \textsc{Ono}\altaffilmark{4}}



\maketitle

\begin{abstract}
The typical accreting neutron star, Aquila X-1, was observed with Suzaku
seven times in the decay phase of an outburst in 2007 September-October.
Among them, the second to the fourth observations were performed
10 to 22 days after the outburst peak,
when the source was in the hard state
with a luminosity of $2\times10^{36}$ erg sec$^{-1}$.
A unified spectral model for this type of objects approximately
reproduced the 0.8--100~keV spectra obtained in these 3 observations.
However, the spectra all exhibited an enigmatic hump-like excess around 30 keV, 
above the hard X-ray continuum which is interpreted as arising via Comptonization.
The excess feature was confirmed to be significant
against statistical and systematic uncertainties.
It was successfully represented by a Gaussian centered at $\sim 32$~keV,
with a width (sigma) of $\sim 6$~keV and an equivalent width of $\sim 8.6$~keV.
Alternatively, the feature can also be explained by a recombination edge model,
that produces a quasi-continuum above an edge energy of $\sim 27$~keV
with an electron temperature of $\sim 11$~keV
and an equivalent width of $\sim 6.3$~keV.
These results are discussed in the context of atomic features of heavy elements
synthesized via rapid-proton capture process during thermonuclear flashes.
\end{abstract}

\section{Introduction}
\label{Introduction}

A Neutron-Star Low-Mass X-ray Binary (NS-LMXB) is
a binary system composed of a Roche-lobe filling low-mass star
(less than a solar mass) and a mass-accreting neutron star.
Because the neutron stars in typical LMXBs are only weakly magnetized, 
the accreting matters are considered to fall mainly onto their equatorial regions, 
which sometimes expand over the entire neutron star surface \citep{Sakurai+2012}. 
There, the matter is compressed, heated, 
and eventually undergoes thermonuclear flashes called Type-I bursts.
These flashes are considered to involve extensive nucleosynthesis,
particularly via rapid-proton-capture process (rp-process)
because of the hydrogen-rich environment (unless the mass donor is a helium star).
The rp-process on accreting neutron stars has been studied theoretically \citep{Schatz+2001}, 
but their observational confirmation is still lacking.
We are hence urged to search LMXB spectra 
for spectral features of the produced heavy elements, in energies of $> 10$~keV
where their K-shell energies fall (e.g., \cite{Strohmayer_2002}).

NS-LMXBs have two spectral states,
called the soft state and the hard state, realized
when the mass accession rate from the donor star is higher and lower,
respectively (White and Mason \yearcite{White_Mason_1985},
Mitsuda and Tanaka \yearcite{Mitsuda_Tanaka_1986},
\cite{Lin+2007}).  
In the soft state, an optically-thick accretion disk is formed 
down to a close vicinity of the neutron star surface,
and its emission makes the system bright
in soft X-rays (typically below 10--20~keV).
Whereas in the hard state, the accretion disk truncates 
at a radius larger than the neutron star surface,
and turns into an optically thin hot flow called a ``corona".
(The corona may still be present in the soft state, with a lower electron temperature and a higher optical depth.)
As this corona strongly Comptonizes blackbody photons 
from the heated neutron-star surface, 
the spectra extend up to $\sim100$~keV
(Sakurai et al. \yearcite{Sakurai+2012}, \yearcite{Sakurai+2014}).

Since the 1980s, NS-LMXB continua have been modeled in various ways 
(e.g. Mitsuda et al. \yearcite{Mitsuda_1984}, Barret \yearcite{Barret_2001}, Lin et al. \yearcite{Lin+2007}, Sakurai et al. \yearcite{Sakurai+2012}, \yearcite{Sakurai+2014}, and Armas et al. \yearcite{Armas_2017}). 
However, our purpose is a study of local spectral features, rather than the modeling of continuum. 
Therefore, for simplicity, we adopt the modeling developed by Sakurai et al. (\yearcite{Sakurai+2012}, \yearcite{Sakurai+2014}). 
According to this modeling, the spectra in both states can be represented by a spectral model 
consisting of an optically thick disk emission and a Comptonized blackbody, 
with the Comptonization much stronger in the hard state. 
This is an extension of the two-component model developed originally for the soft state \citep{Mitsuda_1984}, 
but it can also be regarded as a three-component model (e.g., \cite{Lin+2007}) 
if the seed blackbody and the Comptonizing corona are counted separately. 
Evidently, the data in the hard state are better suited to our purpose
of searching for heavy-element features,
because the emission in $> 10$~keV would be too weak
(relative to that in $< 10$~keV)  if a source is in the soft state.

Aquila X-1 (hereafter Aql X-1) is one of the widely studied NS-LMXBs,
characterized by recurrent outbursts. 
It was observed by the Suzaku satellite \citep{Suzaku}
from 2007 September 28 to October 30, seven times covering an outburst.
This source is suitable for the study of nucleosynthesis processes, 
because it produces Type-I bursts. 
Moreover, it is relatively nearby, 
and usually undergoes the hard state in the rise and decay phases of its outbursts 
which recur with typical intervals of several months to a few years.

\citet{Sakurai+2012} analyzed the spectra of Aql X-1
taken in these Suzaku observations, and constructed the model which we adopt.
Although the spectral fits with that model were statistically acceptable,
a hump structure around 30 keV was seen in the residuals
of the fit to a luminous hard-state spectrum,
as shown in figure 7 in \citet{Sakurai+2012}.
The feature was noted by \citet{Sakurai+2012},
and partially attributed to the reflection signal, but not considered in further details.
In the present paper, we study this hump structure,
through more detailed re-analysis of the same Suzaku data sets
as used by Sakurai et al (\yearcite{Sakurai+2012}, \yearcite{Sakurai+2014}).
We attempt to interpret the feature in the context of
thermonuclear synthesis on the neutron-star surface.

\section{Observation and data reduction}
\label{Observation and data reduction}
As mentioned in section 1, Aql  X-1
was observed with Suzaku during the decay phase
of the outburst in September--October 2007, in seven separate pointing
which are called Obs. 1, Obs. 2, ... and Obs. 7 after \citet{Sakurai+2012}.
Out of these data sets, the present paper utilizes the data of  ObsID =
402053020 (Obs. 2),
402053030 (Obs. 3), and
402053040 (Obs. 4),
acquired on 2007 October 3, October 9, and October 15, respectively. 
These epochs correspond to 10, 16, and 22 days, after the outburst peak 
which was reached on September 23. 
Among them, Obs. 3 was employed in figure 7 of \citet{Sakurai+2012}.
On these occasions, Aql X-1 was in a relatively luminous hard state,
with a 0.8--100~keV luminosity of $\sim2\times10^{36}$~erg sec$^{-1}$.
To these data sets, we applied the same data screening criteria
as used in \citet{Sakurai+2012}.

In theses observations, the X-ray Imaging Spectrometer
(XIS; Koyama et al. \yearcite{Koyama_2007}) onboard Suzaku
was operated with the $1/4$ window mode.
The data were processed with HEASoft (version 6.18)
and the caldb released on 2015 September 14.
We accumulated the on-source events within a circle of $2'$ radius,
and the background events were extracted from a rectangle region next to the source region.
Pileup effects were calculated by the Suzaku pileup tools \citep{Yamada+2012}.
Since the high energy band near 10~keV is important in our analysis,
we need to reduce the pileup effects to less than 1\%.
Therefore, the image center was excluded following the procedure of this tool,
even though \citet{Sakurai+2012} did not conduct the pileup elimination.
In Obs. 2, 3, and 4, the image center was excluded with a radius of
$\sim$20, $\sim$27, and $\sim$18 pixels in the detector coordinates, respectively.
As reported in Sakurai et al. (\yearcite{Sakurai+2012}, \yearcite{Sakurai+2014}),
no X-ray burst was detected in any of these observations, 
and the time variation was less than 10\%.
Therefore,  time-averaged data were utilized for our spectral analysis.
To avoid the instrumental Si K-edge and Au M-edge,
where calibration uncertainties are large,
the 1.7--2.4~keV energy range was excluded.

Like in \citet{Sakurai+2012}, 
the data from the Hard X-ray Detector (HXD; Takahashi et al. \yearcite{Takahashi_2007}) 
were screened with the pipeline tools of $\tt aepipeline$ and $\tt hxdpin(gso)xbpi$ in FTOOLS. 
The Non-X-ray Background (NXB) spectra were created 
from a fake event file provided by the HXD team. 
The Cosmic X-ray Background (CXB) spectrum was simulated by 
the $\tt hxdpinxbpi$ tool using the CXB parameters of \citet{Boldt_1987}.
Then, the NXB and CXB were both subtracted from the on-source spectra.
In the three observations, the HXD-PIN and HXD-GSO signals
were thus detected over 15--50~keV and 50--100~keV, respectively.

\section{Spectral analysis}
\label{Spectral analysis}
In this section, we try to fit the individual spectra from Obs. 2, 3, and 4,
with a canonical LMXB emission model and its variants,
to confirm the presence of the hump-like excess.
Then, in order to quantify it, the three spectra are summed together
to maximize the statistics.
The analysis employs the XSPEC (version 12.9.1) tool,
in which all the spectral models used in the present study are predefined. 

\subsection{Individual Spectra}
\label{individual spectra}
From each of the three observations,
we produced the XIS and HXD (PIN+GSO) spectra,
and fitted them simultaneously,
with the two-component spectral model of NS-LMXBs
constructed by Sakurai et al. (\yearcite{Sakurai+2012}, \yearcite{Sakurai+2014}).
We introduced a factor of 1.158 on the HXD-PIN and HXD-GSO spectra 
to take into account their cross-calibration relative to the XIS \citep{Kokubun_2007}.
As already described briefly in section \ref{Introduction},
the model consists of a multi-color blackbody emission
from an optically-thick disk, represented by {\tt diskbb},
and a single-zone Comptonized blackbody emission model,
represented by {\tt compPS} \citep{Poutanen_Svensson_1996}.
Seed photons of the {\tt compPS} model are assumed to be
the blackbody radiation from the neutron star surface.
A reflection component in {\tt compPS} was taken
into account after \citet{Sakurai+2012}. 
We chose the {\tt compPS} geometry parameter of 4, which means that the corona distributes spherically, 
and fixed the binary inclination to 45$^{\circ}$.
Also we assumed that the coronal electrons follow a Maxwellian distribution. 
Like in the previous studies (\cite{Lin+2007}; \cite{Sakurai+2012}, \yearcite{Sakurai+2014}),
an Fe K$_{\alpha}$ emission line is observed in the present Aql X-1 spectra.
A Gaussian component was therefore incorporated by
fixing its energy and width to 6.4~keV and 0.1~keV, respectively.
The interstellar absorption was modeled by {\tt tbabs} with the Solar abundance \citep{Wilms_2000},
of which the absorption column density was fixed to the same value as
Sakurai (\yearcite{Sakurai+2012}, \yearcite{Sakurai+2014}),
$N_{\rm H}=0.36\times10^{22}$~cm$^{-2}$. 
Here and hereafter, our fits incorporate a systematic error of 1\% for the XIS and HXD-PIN spectra. 
We thus fitted the individual spectra with the
{\tt tbabs}*\{{\tt diskbb} + {\tt compPS(bbody)} + {\tt Gaussian(FeK)}\} model,
which hereafter we refer to as ``the canonical model".

\begin{figure*}[htb]
\begin{center}
\includegraphics[width=17cm,bb=10 180 600 800,clip]{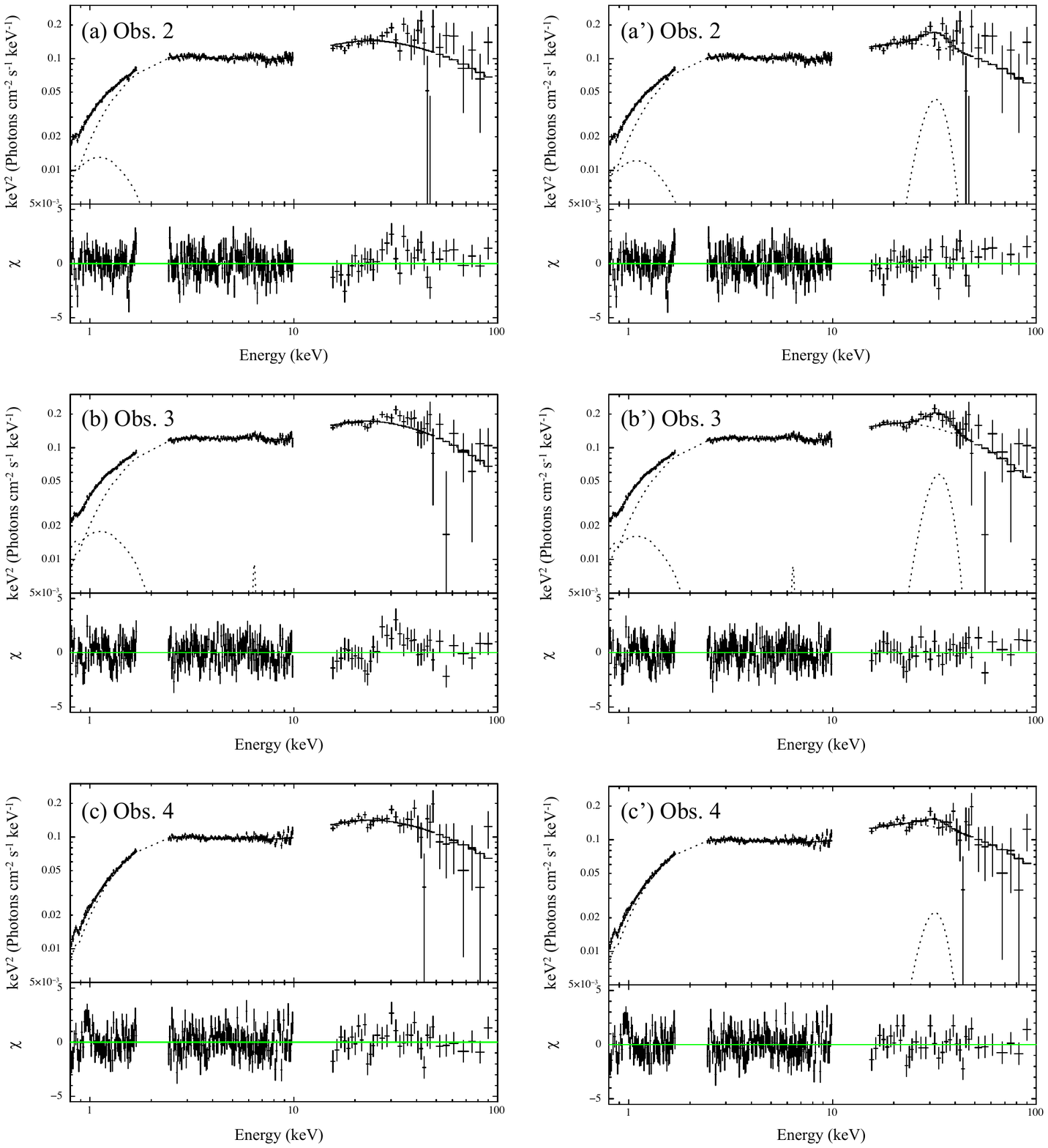}
\end{center}
\caption{
Model fits to the individual Suzaku (XIS+HXD) spectra of Aql X-1.
Panels (a), (b), and (c) are spectra of Obs. 2, 3, and 4,
respectively, fitted by the canonical model.
Panels (a'), (b'), and (c') are the same spectra,
but a Gaussian component was added at $\sim 30$ keV.
}
\label{Individual spectra}
\end{figure*}

Panels (a)--(c) of figure \ref{Individual spectra} shows
the unfolded spectra and model from the three observations, all fitted with the above model.
The reduced chi-squared values, given in table \ref{Gaussian Param}, implies
that the fits are acceptable with null-hypothesis probabilities of $>5\%$.
In addition, three spectra were found to have consistent (within errors)
spectral shape parameters, including
the innermost disk temperature of $kT_{\rm in} \sim 0.24$ keV,
the blackbody seed photons temperature of $kT_{\rm bb} \sim 0.48$ keV,
the Comptonizing electron temperature of $kT_{\rm e} \sim 55$ keV,
the {\tt compPS} optical depth of $\tau \sim 1.1$,
and the  {\tt compPS} reflection solid angle of $\sim 2 \pi$.

Although the canonical model was thus generally successful,
the fit results all exhibit an excess feature  at 30~keV,
which  is most prominent in Obs. 3;
this is the issue to be studied in the present paper.
To examine whether the feature is statistically significant,
and if so, whether it is consistent among the three observations,
we temporarily added a Gaussian component
around  30 keV to the canonical continuum.
In the fits to the Obs. 2 and Obs. 4 spectra,
the Gaussian width, $\sigma$, was not well constrained;
so it was fixed to 4.5 keV which was obtained from Obs.3.
The obtained fits are presented in panels (a')--(c') of figure \ref{Individual spectra},
and the best-fit parameters of the Gaussian component
are listed in table \ref{Gaussian Param}.
The derived lower limits on the equivalent widths (EW) are all positive (at 90\% confidence level), 
implying that the excess is significant.
Furthermore, within errors, the EW is mutually consistent among the three spectra,
and so is the Gaussian centroid energy $E_{\rm c}$.
The continuum parameters did not change significantly when adding the Gaussian component.
Thus, all the three spectra significantly and consistently exhibit
the hump-like excess around 30 keV above the {\tt compPS} continuum. 

\begin{table*}[h]
 \caption{Gaussian parameters obtained from the individual spectra.
 The symbols are defined in text.}
 \label{Gaussian Param}
  \begin{center}
  \setlength{\tabcolsep}{8pt}
  \normalsize
  \begin{tabular}{lccccc}
 \hline \hline 
   Obs. ID         &  $E_{\rm c}$       &    $\sigma$            &    EW       &   \multicolumn{2}{c}{$\chi_{\nu}^2 (\nu)$} \\
                        &      (keV)              &    (keV)                  &  (keV)      &   w/o Gaussian    &  w Gaussian \\
  \hline 
  Obs. 2           &  $31_{-3}^{+4}$  &   4.5 (fixed)              & $7.3_{-3.2}^{+3.2}$  & 1.13 (219)  &  1.08 (217) \\
  Obs. 3           &  $33_{-2}^{+3}$  &$4.5_{-2.1}^{+4.1}$  & $8.8_{-3.5}^{+6.4}$  & 1.00 (219)  &  0.88 (216)   \\
  Obs. 4           &  $31\pm5$          &   4.5 (fixed)              & $3.7_{-2.5}^{+2.6}$  & 1.22 (219)  &  1.21 (217)  \\
\hline 
  \end{tabular} \label{Gaussian Param}
 \end{center}
\end{table*}

\subsection{Analysis of the merged spectrum with different continuum models}
Now that the three spectra were confirmed in subsection \ref{individual spectra}
to exhibit a consistent continuum shape and a consistent  30~keV hump,
they were summed up together with
{\tt addascaspec}  in HEASoft to improve the statistics.
To the GSO data points, we added (in quadrature) a large systematic error, 
that is, $1\sigma$ systematic uncertainty of NXB, which is comparable to $\sim 20\%$ of the signal. 
For reference, this systematic error of the GSO data were negligible when analyzing the individual spectra, 
due to a larger statistical errors.

\subsubsection{Fit with the canonical continuum model}
\label{Fit with the canonical model}

To the merged spectrum, we applied the same canonical model as in subsection
\ref{individual spectra}.
Then, as shown in figure \ref{Fit Spectrum}a,
the spectrum was roughly represented over the 0.8--100 keV range by this model,
and the obtained parameters were all consistent 
with those obtained from the individual spectra in subsection~\ref{individual spectra}.
However, the positive residual is again seen around 30~keV,
even though the fit considered the reflection effect.
Because of the improved statistics, the fit has worsened to $\chi_{\nu}^2~(\nu) =$1.23 (219), 
implying a null-hypothesis probability of 1.23\% which is regarded as not yet satisfactory. 
To confirm that the fit failure is due to the 30~keV excess,
the same fit was repeated by ignoring the data in the 23--40~keV energy range.
The fit quality was then significantly improved to $\chi_{\nu}^2~(\nu)=$1.04 (205).
This result shows that the canonical model is rejected
due to the 30~keV hump.

For reference, the feature remained similarly significant, 
even when we employed other continuum models with reflection effects, 
including {\tt relxill}  \citep{Magdziarz_1995} or {\tt pexriv} \citep{Garca_2014}. 
This is understandable, because these reflection models are all based on Compton scattering and photoelectric absorption in thick neutral matter, 
and would not differ significantly in local spectral shapes (except possibly the iron-K edge). 


\subsubsection{Fit with a modified canonical model}

The canonical model has several variants,
including the one adopted by  \citet{Lin+2007};
it assumes that a fraction of blackbody photons from the neutron-star surface are Comptonized, 
whereas the rest reach us directly without Comptonized.
We hence constructed the model as {\tt tbabs*\{diskbb + bbodyrad + compPS(BB) + Gaussian(FeK) \}},
where  {\tt bbodyrad} represents the directly-visible blackbody component
with its normalization  left free
and its temperature  tied to the seed-photon temperature of  {\tt compPS}. 
However,  when this model was applied to the merged spectrum,
the normalization of  {\tt bbodyrad} 
became  consistent with zero within errors,
and the other parameters were essentially the same  as those obtained in subsubsection \ref{Fit with the canonical model},
including the blackbody temperature of  $0.48 \pm 0.01$~keV.
The radius (assuming a spherical emission region) was constrained as $< 2.3$~km (90\% limit).
Therefore, this model has essentially reduced to the canonical model. 
Naturally, the fit quality did not improve, with $\chi^2(\nu) = 1.23~(218)$.

\subsubsection{Double Comptonization models}
\label{Double Comptonization model}
One of possible explanations of  the 30~keV hump is
to regard it as an artifact that arises because the continuum models considered so far are too simple.
In particular, the Comptonizing corona can have more than one electron temperatures,
so that  the different cutoff energies and optical depths
will produce somewhat different continuum shapes.
We are hence motivated to test ``double Comptonization" modelings,
in the following two configurations.

One idea is to assume that the blackbody emission from the neutron
star surface is Comptonized by a corona surrounding the neutron star,
whereas the emission from  the disk is partially Comptonized
by a different corona (possibly localized on the disk surface).
Actually, this modeling was employed by \citet{Sugizaki+2013}
and \citet{Sakurai_2015} to avoid deficits in the seed photons,
and by \citet{Ono+2016} to reproduce the 25--70 keV part of the spectrum of GS 1826$-$238.
To examine this form of disk Comptonizaton,
we replaced  the {\tt diskbb} component of the model in
section \ref{Fit with the canonical model}
with a  {\tt dkbbfth} model \citep{Done_Kubota_2006},
in which the disk is assumed to be covered by a Compton corona
from the innermost radius $r_{\rm in}$  up to a lager radius $r_{\rm tran}$.
The model has five parameters;
$kT_{\rm in}$,  $r_{\rm tran}$, $kT_{\rm e}$,
the photon index of the Comptonized component which is related to $\tau$,
and the normalization which is translated into $r_{\rm in}$.

The merged spectrum fitted by this double Comptonization model
is shown in figure \ref{Fit Spectrum}c.
The fit improved to  $\chi_{\nu}^2~(\nu)=$1.11 (216),
which is acceptable with a null-hypothesis probability of 13.6\%.
This fit  improvement has been caused by the following reason.
The disk emission became stronger in the 2--8~keV range 
due to the newly invoked Comptonization, 
and the temperature of {\tt compPS} increased. 
As a result,  the slight negative residuals previously seen at 8--25~keV diminished. 
However,  through equation (A.1) in \citet{Kubota_Makishima_2004},
the fit  gives $r_{\rm in} \lesssim 10$ km,
because a larger value would over-predict the emission below a few~keV.
This disagrees with the fact  that the disk in the hard state
is truncated at a radius of several tens km
(\cite{Ono+2017}, \cite{Sakurai+2012}, \cite{Sakurai+2014}). 
Thus, the model is acceptable in a statistical sense, but physically unacceptable.

The other modeling is simply to assume that the disk is not Comptonized,
whereas the blackbody from the neutron star is Comptonized
by two different coronae with different temperatures.
We employed two {\tt CompPS} components
to represent this form of double-Comptonization, and constructed a model as
{\tt tbabs*\{diskbb + compPS(bbody) + compPS(bbody) + Gaussian(FeK)\}}.
The two {\tt CompPS} components were constrained
to have the same seed-photon temperature,
but allowed to take separate and free $kT_{\rm e}$ and $\tau$.
As presented in figure \ref{Fit Spectrum}d, this model gave $\chi_{\nu}^2(\nu) = 1.12~(216)$ when applied to the overall 0.8--100.0~keV energy range.
To examine the fit goodness around 30~keV, 
we again fitted the spectrum only using the energy range of 8.0--100.0~keV,
fixing $kT_{\rm in}$ and the reflection-component intensity
to the values obtained from the 0.8--100 keV fit.
As a result, the fit has become statistically unacceptable
with $\chi_{\nu}^2(\nu) = 1.56~(44)$ 
with a null-hypothesis probability of $\sim 1.1\%$.
Thus, the 30 keV structure cannot be explained away
by the 2nd double-Comptonization modeling, 
either;  the favorable $\chi_{\nu}^2$ in the 0.8--100~keV 
fit is mainly due to ``dilution" by the large number of degree of freedom in the 0.8--8~keV range. 
We do not consider this case hereafter. 

\subsubsection{Partial covering model}
\label{Partial covering model}

Another possible modification of the continuum model to explain
the 30~keV hump is to assume partial covering configuration
(e.g., \cite{Iaria_2013}). 
It represents a condition wherein
emission from a source is partially covered by a thick absorber,
parametrized by its hydrogen column density $N_{\rm H} '$
and a covering fraction.
Since the spectrum in this case is thus the sum
of a directly observed emission (without the extra absorption)
and the absorbed one, it can take more complex shapes,
depending on $N_{\rm H}'$ and the covering fraction.

Along the above consideration, we multiplied the {\tt pcfabs} factor 
to the canonical model, to construct yet another modified model as 
{\tt tbabs*pcfabs*\{diskbb + compPS(bbody) + Gaussian(FeK)\}}. 
The fit result by this model is shown in figure \ref{Fit Spectrum}e. 
Thus, the spectrum around 30~keV was considerably better reproduced, 
with the fit goodness of $\chi_{\nu}^2(\nu) = 1.07~(217)$.
However, the 30~keV excess is still visible. 
Furthermore, the best-fit $N_{\rm H}'$ became 
very high as $1.6 \times 10^{25}$~cm$^{-2}$, 
and the corona covered a significant fraction ($\sim 0.3$) of the disk. 
If such a dense absorber, that has a Compton optical depth $>1$
and a significant covering fraction,
were located between the observer and the continuum source,
X-rays from the source would be scattered into $4\pi$ directions,
and hence a very high luminosity would be required.
Therefore, the partial covering modeling is also unphysical;
we do not discuss it any further, either.

\subsection{Modeling of the 30 keV feature}

Since the 30~keV hump was not accounted for by trimming the continuum model,
we regard it as real, as long as we consider only the statistical data uncertainties
(systematic effects to be evaluated later).
Then,  the next step would be to add, on top of the canonical continuum model,
a spectral component that describes the local feature.

\subsubsection{Gaussian model}
\label{Gaussian model}

The simplest form to express the local excess feature will be a Gaussian.
Although it is primarily empirical,
a Gaussian can have some physical meaning as well,
because it can represent an atomic emission line possibly arising
from some heavy elements that may have been produced in Type-I bursts.
Therefore, we fitted the data with the
{\tt tbabs}*\{{\tt diskbb} + {\tt compPS(bbody)} + {\tt Gaussian(FeK)} + {\tt Gaussian}\} model.

The obtained fit results are shown in figure \ref{Fit Spectrum}f,
and the best-fit parameters are given in table \ref{Fit Parameters}.
Thus, the model has successfully reproduced the spectrum
with the fit goodness of $\chi_{\nu}^2 (\nu)=1.05~(216)$,
which improves over the canonical-model fit
by $\Delta \chi^2 = 42.9$ ($\Delta\nu = -3$).
The associated $F$-value of 13.6 indicates
that the probability for this improvement to occur by chance
is $3.4 \times 10^{-8}$.
The Gaussian is indeed centered at $E_{\rm c} \sim 32$ keV,
and is inferred to be moderately extended by $\sigma/E_{\rm c}\sim 0.2$.
The Gaussian normalization is securely positive,
and translates to an EW of the feature as 8.6$_{-3.2}^{+7.2}$~keV.
These parameter values are consistent with those obtained from the individual spectra (table \ref{Fit Parameters}). 
The continuum parameters did not change significantly
even when adding the Gaussian component.

\begin{table*}[htb]
\tbl{Parameters of the Gaussian and {\tt redge} components.\footnotemark[$*$]}{%
\begin{tabular}{lccc}
\hline
   \multicolumn{1}{c}{Component} & Parameter & Gaussian model & {\tt redge} model \\
\hline
 {\tt diskbb}
           & $kT_{\rm in}$ (keV)      & $0.23\pm0.02$            & $0.23\pm0.02$     \\
           & $r_{\rm in}$ (km) \footnotemark[$\dag$]
                                                    & $33 \pm 7$                  & $33 \pm 7$     \\
 {\tt CompPS}
           & $kT_{\rm e}$ (keV)        & $49\pm4$                    & $49\pm4$     \\
           & $kT_{\rm bb}$ (keV)      & $0.46\pm0.02$            & $0.46_{-0.01}^{+0.02}$     \\
           & $\tau$                            & $1.2\pm0.1$                & $1.2\pm0.1$     \\
           & reflection ($2\pi$)          & $0.93_{-0.23}^{+0.22}$        & $1.0\pm0.2$     \\
           & $R_{\rm{bb}}$ (km)\footnotemark[$\dag$]
                                                     & $12\pm1$                   & $12\pm1$     \\
 {\tt Gaussian}
          & $E_{\rm c}$ (keV)           & $32_{-3}^{+2}$ & -\\
          & $\sigma$ (keV)               & $6_{-2}^{+4}$ & -\\
          & norm ($10^{-4}$)             & $4.8_{-1.8}^{+4.0}$ & - \\
 {\tt redge}
         &$E_{\rm e}$ (keV)             & -                           & $27\pm1$\\
         & $kT_{\rm e}'$ (keV)           & -                           & $11_{-5}^{+10}$  \\
         & norm ($10^{-4}$)               & -                           & $5.0_{-1.7}^{+2.6}$\\
 {\tt Gaussian (FeK)}
         & norm ($10^{-5}$)              & $2.5\pm1.8$       & $2.3\pm1.8$     \\
 Fit goodness
         & $\chi_{\nu}^2~(\nu)$        &  1.05 (216)    &  1.02 (216)   \\
\hline
\end{tabular}}\label{Fit Parameters}
\begin{tabnote}
   \footnotemark[$*$] Errors represent 90\% confidence limits. Symbols are defined in text. \\
   \footnotemark[$\dag$] Calculated assuming a source distance of 5.2~kpc and an inclination angle of 45$^{\circ}$.
\end{tabnote}
\end{table*}

\subsubsection{Absorption edge and recombination edge emission models}
\label{Recombination edge emission model}
If the 30~keV structure is related to some heavy elements,
we may also consider a possibility that it is due to their K-edge absorption.
Therefore, we multiplied an {\tt edge} factor to the canonical model,
to construct a model as {\tt tbabs}*{\tt edge}*\{{\tt diskbb} + {\tt copmPS(bbody)} + {\tt Gaussian (FeK)}\}.
The K-edge energy was varied over the range of 25-50~keV.
However, the fit has remained relatively poor,
$\chi_{\nu}^2 (\nu)=1.24~(217)$,
and the shape of the residuals did not change.

\begin{figure*}[htb]
\begin{center}
\includegraphics[width=15cm,bb=0 0 600 900,clip]{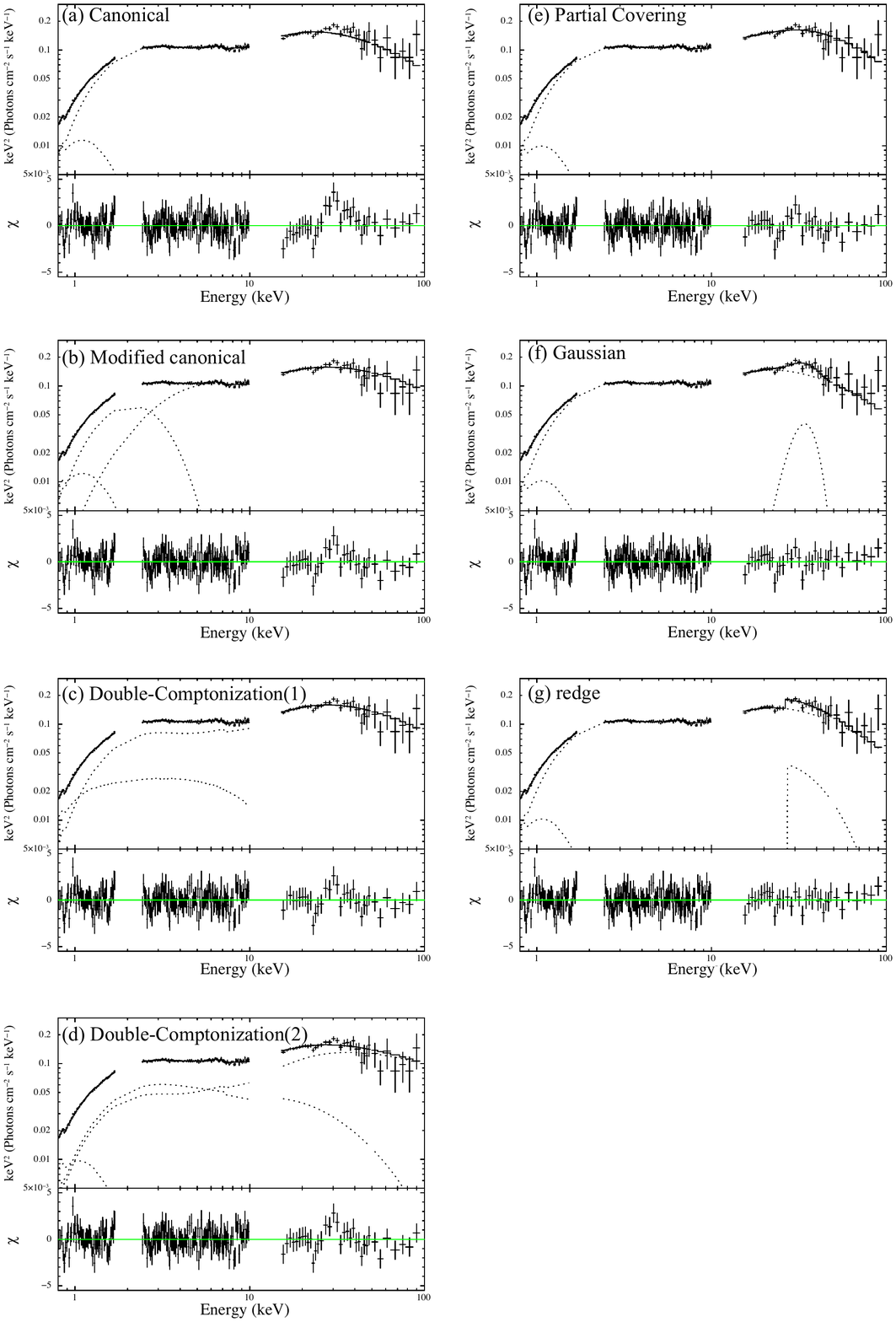}
\end{center}
\caption{
Simultaneous fittings to the summed XIS, HXD-PIN, and HXD-GSO spectra,
presented  in the $\nu F_{\nu}$ form.
(a) A fit with the canonical model.
(b) A fit with the modified canonical model. 
(c) When the double Comptonization model
(the 1st condition described in text)~is employed.
(d) The same as (b), but with the 2nd form of double Comptonization.  
(e) Results with the partial covering model.
(f) The case of adding a Gaussian component.
(g) When an {\tt redge} factor is multiplied to the canonical continuum model.
}
\label{Fit Spectrum}
\end{figure*}

Another possible spectral feature related to the heavy elements
is a recombination edge structure, namely,
a quasi-continuum emission above a K-edge energy.
It is produced when a plasma is strongly photo-ionized by high-energy photons, 
so that heavy ions achieve an ionization temperature 
which is much higher than the kinetic plasma temperature, 
and free electrons in the plasma recombine 
with the ions through free-bound transitions.
The produced quasi continuum is represented by an additive model called {\tt redge},
which in turn is parametrized by the K-edge energy $E_{\rm e}$,
the electron temperature $kT_{\rm e}'$
describing the high-energy extension of the feature,
and normalization.
We fitted the spectrum with the 
{\tt tbabs}*\{{\tt diskbb} + {\tt compPS(bbody)} + {\tt Gaussian(FeK)} + {\tt redge}\} model,
and obtained the results shown in figure \ref{Fit Spectrum}g and table \ref{Fit Parameters}.
The spectrum was successfully explained by this model,
with $\chi_{\nu}^2 (\nu)=1.02~(216)$ which is even better
(by $\Delta\chi^2 = -5.4$) than that of the Gaussian model.
The fit yielded $E_{\rm  e} \sim 27 $~keV and $kT_{\rm e}' \sim 11$~keV, respectively.
Since the latter is considerably lower than the temperature ($\sim kT_{\rm e}$ in table 2) 
of the illuminating hard X-rays, the recombination-edge interpretation is self-consistent. 
Other parameters are listed in table \ref{Fit Parameters}.
Again, the continuum parameters did not change by including the {\tt redge} component.

\section{Discussion} \label{sec.4}
We analyzed three Suzaku spectra of Aql X-1 acquired during the decay phase of the outburst
in 2007 September--October, and confirm that they show a statistically significant hump structure at $\sim30$~keV.
It was successfully represented by a Gaussian centered at $E_{\rm c} = 32_{-3}^{+2}$~keV,
with $\sigma =$ 4--10~keV and an EW of $8.6_{-3.2}^{+7.2}$~keV.
Alternatively, the feature can also be explained by the {\tt redge} model,
with $ E_{e} = 27 \pm 1$~keV, $kT_{\rm e} = $ 6--21~keV, and an EW of $6.3_{-2.2}^{+3.3}$~keV.
Now we discuss the implications of these results.

\subsection{Evaluation of systematic errors}
\label{Evaluate the systematic error}
Before discussing the origin of the 30~keV hump,
we need to evaluate systematic uncertainties of our analysis,
to confirm that the hump structure is not an artifact.
Here, the detector response and the background modeling of HXD-PIN are major sources of the uncertainties.

In order to examine whether the employed HXD-PIN response is accurate enough,
the data of the Crab Nebula obtained on a similar epoch, 
namely, 2008 8 27 (Obs. ID = 103007010) for net 33 ks,  were analyzed using the same response.
The 15--50 keV Crab spectrum taken with HXD-PIN was 
successfully ($\chi^2/\nu = 90.1/90$) represented by a single power-law
of photon index 2.12 $\pm$ 0.01, and the model-to-data ratio in 20--40~keV remained within $\pm3.5$\% of unity.
This is much smaller than the 30~keV hump structure,
which amounts to $\sim40$\% of the 20--40~keV continuum
even employing the modified continuum models (figure \ref{Fit Spectrum}b to figure \ref{Fit Spectrum}d).
Therefore, the observed 30~keV hump structure cannot be an artifact arising
from uncertainties or inaccuracies of the instrumental response.

\begin{figure}[h]
\begin{center}
\includegraphics[width=10cm,bb=100 0 840 595,clip]{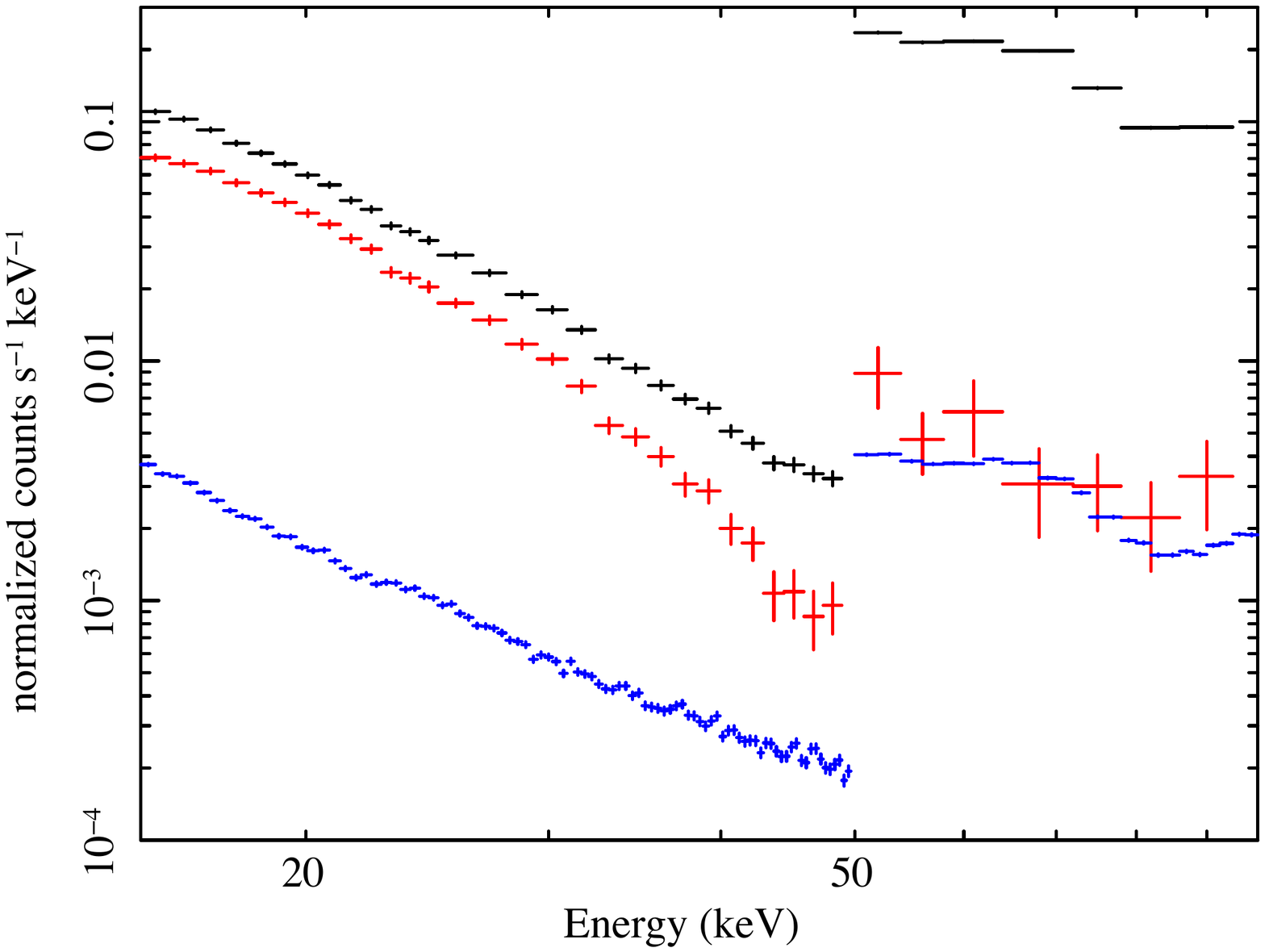}
\end{center}
\caption{
A comparison of the HXD (PIN and GSO) spectra with the 3$\sigma$ NXB uncertainty (blue). 
Black shows the total spectra from the three observations summed up, 
and red are those after subtracting the NXB and CXB. 
The GSO errors include (in quadrature) $1\sigma$ NXB uncertainty. 
}
\label{bg spectrum}
\end{figure}

The NXB component of HXD-PIN is almost featureless, except for the weak Gd-K line at 43~keV
arising from fluorescence in the HXD-GSO scintillators underneath HXD-PIN \citep{Kokubun_2007};
any local features in the 20--40~keV energy band are known to be less than 20\% of the average value.
Since the NXB intensity is about 67\% of the signal from Aql X-1 in the 20--40~keV band,
possible $<20\%$ local features in the NXB component would correspond to those of $<13\%$ in the Aql X-1 spectrum.
This is far insufficient to explain the observed 30~keV feature as described above.
The NXB model for HXD-PIN itself is known to contain 3\% systematic uncertainty \citep{Fukazawa_2009},
but this cannot explain the $\sim$40\% excess, either.
To visualized these conditions, we compare in figure \ref{bg spectrum} the signal spectra with 3$\sigma$ 
(i.e., a chance probability of $\sim 10^{-3}$) NXB uncertainties. 
From these evaluations, we conclude that the 30~keV hump structure is significant
even considering both the statistical and systematic errors involved in the data.

Incidentally, the GSO signal in figure \ref{bg spectrum} is comparable to the $3\sigma$ NXB uncertainty. 
However, the detection is significant at 90\% ($1.28 \sigma$) confidence level which is a standard criterion. 
We have hence retained the GSO data, incorporating the $1\sigma$ NXB uncertainty (section \ref{individual spectra}).

\subsection{Is the structure universal?}
\label{Is the structure universal?}
To further confirm that  the feature is not due to  Suzaku-specific
artifacts other than the two considered above,
we searched the literature for similar phenomena.
Then, a 3--120 keV spectrum of Aql X-1,
acquired  with the PCA+HEXTE onboard RXTE in the hard state during a small outburst 
in 2004 February \citep{Lin+2007}, 
was found to exhibits a similar excess structure localized at 30--34 keV (their figure 5).
Actually, this effect apparently motivated these authors to fit
the hard X-ray part of their spectrum with a broken power-law with a break at $\sim 30$~keV.
Furthermore, a similar local excess feature at 30--40~keV, though much weaker and less significant, 
could be present in a joint INTEGRAL+BeppoSAX spectrum
of the NS-LMXB, 4U~1812-12 \citep{Tarana+2006}.
On the other hand, such a spectral structure is apparently absent
in many other published spectra of LMXBs in the hard state,
including those from Aql X-1 itself \citep{Rodriguez_2006}
and other LMXBs such as 4U 1705$-$44 \citep{Lin+2010}
and 4U~1608$-$52 \citep{Armas_2017}.
We hence arrive at two important suggestions.
One is that the feature is neither a Suzaku artifact,
nor specific to Aql X-1.
The other is that it is likely to be visible
only under some limited conditions of NS-LMXBs.
These inferences are qualitatively consistent with our standpoint of
regarding the phenomenon as atomic features of some heavy elements,
because they would be produced occasionally
in Type I bursts near the neutron-star surface,
and would become invisible on a certain timescale as they are
buried beneath the fresh metal-poor accreting materials.

Under which conditions, then, is the feature detectable?
Leaving the study of other sources to a separate publication,
let us focus here on Suzaku data of Aql X-1,
and look at the remaining four data sets  from the 2007 outburst.
From Obs. 1, however, no useful constraint is available,
because the object was then in the soft state,
and hence the intensity at $\sim 30$ keV was  an order of magnitude lower
than in the three observations analyzed here (figure 2 of Sakurai et al \yearcite{Sakurai+2014}).
The same is the case with Obs. 5,
when the source was in the hard state but
was about an order of magnitude dimmer than in the three.
Finally,  the signal was undetectable with the HXD
in Obs. 6 and Obs. 7 when the source was faintest.
Thus, no useful information is available
from these 4 additional data sets covering the 2007 outburst.

\begin{figure}[h]
\begin{center}
\includegraphics[width=10cm,bb=90 0 840 595,clip]{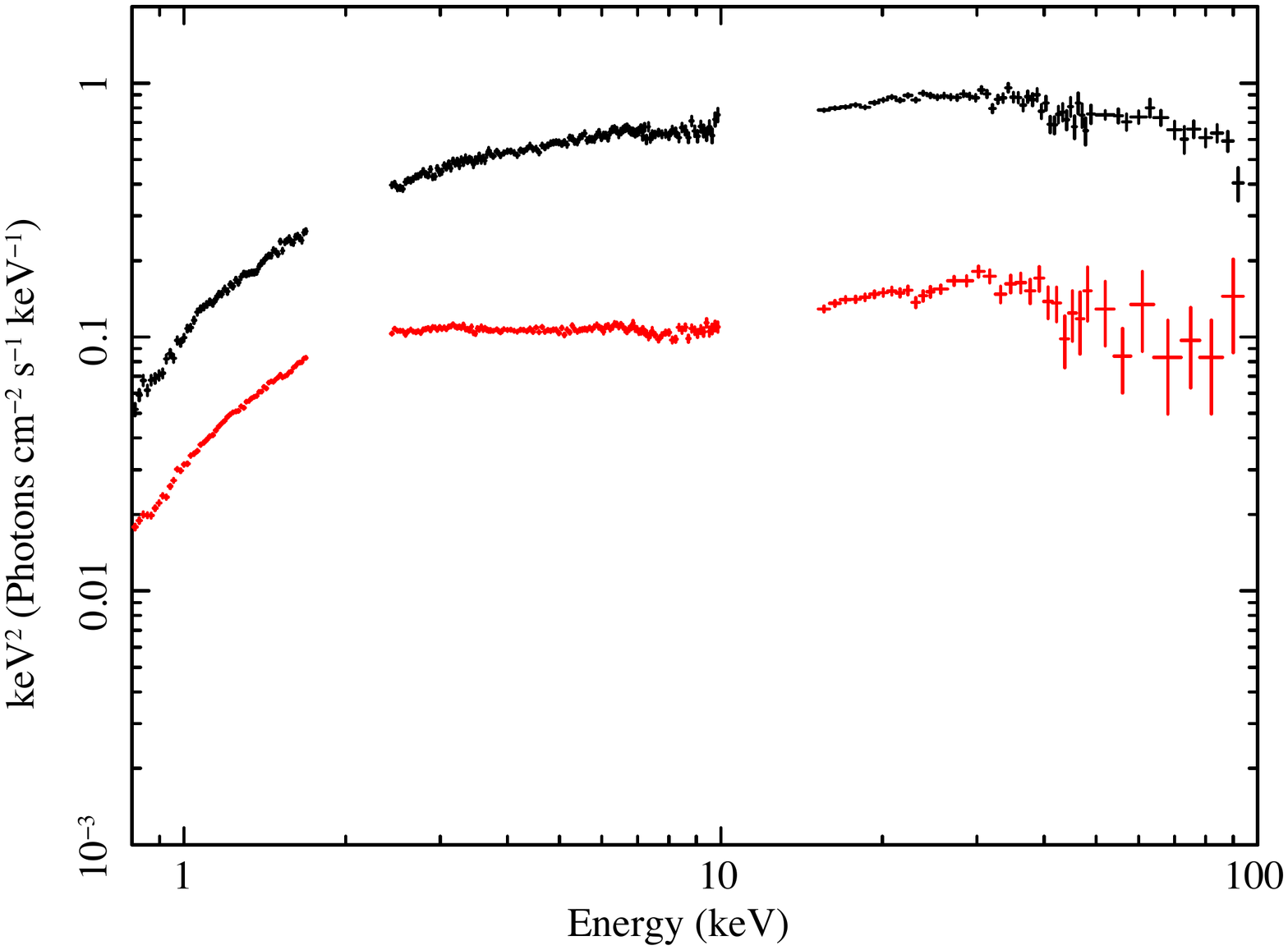}
\end{center}
\caption{
A comparison of two spectra (in $\nu F_{\nu}$ form) of Aql X-1 obtained by Suzaku. 
Red shows the merged data in 2007. 
Black shows a spectrum taken in 2011 October 18, 
from UT03:42:33 to 2011 from October 19 02:39:18, 
just before the hard-to-soft transition which was reported by \citet{Ono+2017}.
}
\label{2011 wo spectrum}
\end{figure}

\begin{figure*}[h]
\begin{center}
\includegraphics[width=15cm,bb=10 100 820 430,clip]{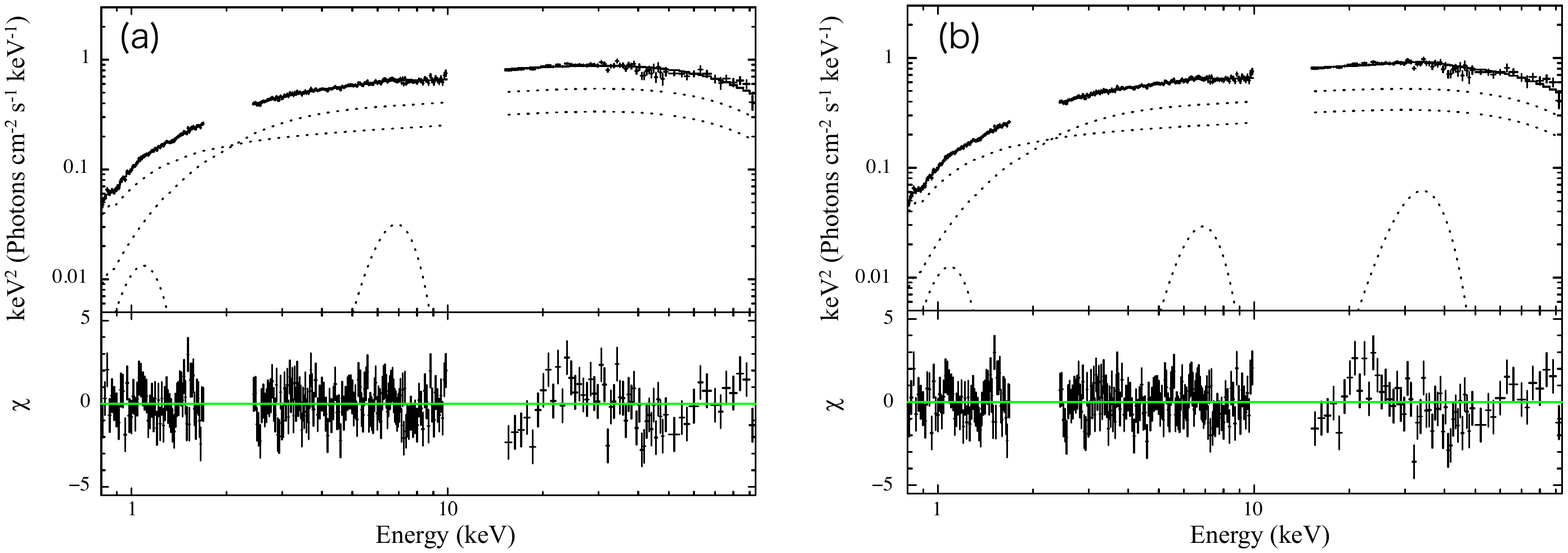}
\end{center}
\caption{
Simultaneous fitting to the XIS 0, HXD-PIN, and HXD-GSO spectra in 2011 presented in figure \ref{2011 wo spectrum}.
(a) A fit with the {\tt tbabs*\{nthcomp(bbody) + nthcomp(disk) + Gauss(1keV) + Gauss(6.6keV)\}} model. 
(b) The model is basically the same as (a) but a Gaussian model was added at 32~keV. See text for details. 
}
\label{2011}
\end{figure*}

Aquila X-1 was in fact observed  with Suzaku again,
in a rising phase of another outburst which took place in 2011 October.
As already published by \citet{Ono+2017},
the observation lasting for one day caught
the source at first in a luminous hard state,
and then witnessed a remarkable hard-to-soft state transition.
However, the spectrum ``P0" of  \citet{Ono+2017},
a hard-state data set obtained from a pre-transition period,
does not show such a hump feature at $\sim 30$ keV.
To further examine this inference, 
we accumulated the data over a longer time (32.8 ks net exposure) 
before the transition, and obtained the spectrum shown in figure \ref{2011 wo spectrum}. 
The merged 2007 spectrum is also shown there for reference. 
Thus, the spectrum in 2011 is $\sim 5$ times brighter than in 2007, 
with a considerably harder Comptonization slope (e.g. in 3--20~keV), 
and shows no particular structure at $\sim 30$~keV in agreement with \citet{Ono+2017}. 
To fit the spectrum we employed a double-seed continuum model as
{\tt tbabs}*\{{\tt nthcomp(diskbb)} + {\tt nthcomp(bbody)} + {\tt Gaussian(1keV)} + {\tt Gaussian(FeK) + {\tt Gaussian(32keV)}}\} model,
after Sakurai (2015), who found that the canonical single-zone
Comptonization model suffers from a shortage in the seed photon flux.
Because the harder continuum suggests a relatively high values of $\tau$, 
we replaced {\tt compPS}, which is valid for $\tau < 3$, with another model, 
{\tt nthcomp} \citep{Zdziarski+1996}, which can be used for $\tau > 2$. 
Over the overlapping range of $\tau = 2$ to 3, the two codes are known to give consistent results \citep{Sakurai_2015}. 
Figure \ref{2011}a shows the fit with this model, 
with a goodness of $\chi_\nu^2(\nu)=1.24~(240)$. 
The {\tt nthcomp} parameters were obtained as $kT_{\rm e} = 20 \pm 1$~keV, 
$kT_{\rm bb} = 0.56 \pm 0.03$~keV (seed photons), and $\tau = 3.67 \pm 0.02$ 
indicated in figure \ref{2011 wo spectrum} by a harder continuum shape, 
as calculated from the photon index describing the model. 
Thus, the choice of {\tt nthcomp} is self consistent 
because the model works only for $\tau > 3$. 
By further adding a Gaussian, with $E_{\rm c}=32$ keV
and $\sigma=5.9$ keV both fixed to the value in table 2,
its EW was constrained as $1.22_{-0.78}^{+0.80}$ keV (90\% confidence).
Although the zero EW is still excluded, the upper limit of 2.02 keV
is lower than the lower bound in the 2007 data, 4.9 keV.
Therefore, the hump feature is significantly weaker (in EW)
than in the merged spectrum from the 2007 outburst.
Figure \ref{2011}b shows the fit when the Gaussian is forced
to take the allowed maximum EW,
and the other parameters are all re-adjusted.

How can we explain the above difference between the 2007 and 2011 data?
The simplest possibility would be that the source was burst active in 2007,
and inactive in 2011.
However, this is rather opposite;
the seven observations in 2007 caught no Type I bursts \citep{Sakurai+2014},
whereas the 2011 pointing recorded at least nine Type I bursts
in a net exposure of about 30 ks.
An alternative explanation may be as follows.
In the merged 2007 spectrum,
the Comptonizing corona was relatively thin with $\tau \sim 1$ (table \ref{Fit Parameters}),
but in 2011 when the object was more luminous,
the corona was rather optically thick with  $\tau = 3.67 \pm 0.02$ as found with figure \ref{2011}.
Then in 2011, any local spectral feature
that arises from a vicinity of the neutron-star surface would have become
undetectable due to smearing through the  repeated Compton scattering,
This interpretation, if correct, supports the view 
that the feature is actually produced at or near the neutron-star surface.

\subsection{Interpretations of the 30 keV structure as an atomic feature}
\label{Interpretation of the 30 keV structure}

\subsubsection{General consideration}
From the analysis in subsection 3.2,
the 30 keV bump may be interpreted
as a K-shell feature of  some heavy ions.
To identify the appropriate elements,
let us recall the Moseley's law (e.g., \cite{Hohenemser_1968}),
which approximates the K-edge energy of ions of atomic number $Z$ as
\begin{equation}
E_{\rm K} \approx (1+g)^{-1}  E_0 (Z-\delta)^{2}~
= 33 \left[ (Z-\delta)/55 \right]^{2} ~\rm({keV})~.
\label{eq:Moseley}
\end{equation}
Here, $E_0 = 0.0136$ keV is the ionization potential of Hydrogen atoms,
$g$ is the gravitational redshift on the neutron-star surface
which we take as $g=0.23$ (assuming a mass of 1.4 $M_\odot$ and a radius of 12 km),
and $\delta$ is a correction factor representing the ionization state;
$\delta=0$ for Hydrogen-like ions, and $\delta \sim 1$ for neutral ones.
The K$_\alpha$ line energy is approximately $0.75 E_{\rm K}$.
From this scaling,  the relevant  elements are
estimated to have $Z \sim 50-60.$

If referring to the Solar abundances, 
the heavy elements with such high $Z$ 
would be extremely scarce, 
$\lesssim 2 \times10^{-10}$ by number relative to Hydrogen,
or $\lesssim 5 \times10^{-6}$  relative to Iron.
However, as mentioned in section 1,
elements up to these atomic numbers can be synthesized
via Type I bursts involving the rp-process \citep{Schatz+2001}.
These heavy elements,
produced somewhat below the neutron-star surface,
may be dredged up into the atmosphere,
possibly via convection or other processes.
There, the heavy-element atoms are subject to two
ionization/excitation processes.
One is bombardment by protons and electrons in the corona,
which are falling onto the atmosphere at a speed
which is a fraction of the free-fall velocity.
The other is irradiation by the hard X-ray photons,
which are produced when the outgoing blackbody soft X-rays
are Compton-scattered back by the coronal electrons.

As a consequence of the above  processes,
the heavy-element atoms in the atmosphere
will attain a relatively high ionization states,
and produce two types of characteristic X-rays.
One is emission of K-line (particularly K$_\alpha$) photons,
which occurs when a remaining K-shell electron is
collisionally excited into a higher bound energy level.
The Gaussian modeling descried in subsubsection~\ref{Gaussian model}
applies to this condition.
The other is production of a recombination quasi-continuum above $E_{\rm K}$.
As already touched on at subsubsection \ref{Recombination edge emission model}
in our redge modeling,
this process mainly occurs when the heavy ions undergo 
K-shell  photo-ionization by  the hard X-ray photons,
followed by recombination with ambient electrons. 
Such free electrons would be abundant in the atmosphere,
because Hydrogen and Helium therein must be
almost completely ionized through irradiation by
the blackbody photons with a sub-keV temperature arising from the neutron-star surface.
Below, we try to identify the corresponding elements
from the Gaussian and redge modelings of the 30 keV feature.

\subsubsection{Gaussian model}\label{Gaussian model(D)}
In figure \ref{Energy vs. atomic num.},
the solid line and the open squares show
the K$_{\alpha}$ line energies of heavy elements as a function of $Z$,
calculated more accurately than with equation \ref{eq:Moseley} 
which is rather approximate, and incorporating $g=0.23$ assuming that the feature arise on the neutron star.
Only the neutral and H-like conditions are shown,
because the other ionization states fall in between them.
There, the neutral K$_{\alpha}$ energies refer to the  weighted average
of those for K$_{\alpha 1}$ and K$_{\alpha 2}$,
as  taken from  the X-ray booklet \citep{XrayBooklet}.
In the case of H-like ionization state,
the K$_{\alpha 1 }$ and K$_{\alpha 2}$ energies
were taken from the AtomDB data base (version 3.0.8),
and the same relative intensity as the neutral atoms was assumed.
Because the AtomDB provides the data only up to Kr ($Z=36$), 
the data from $Z=2$ to $36$ were fitted by the equation as
\begin{equation}
E_{\rm K} (Z) = a\left(Z - \delta \right)^c ~{\rm eV}
\end{equation}
which is essentially the same as equation (\ref{eq:Moseley}), 
but the parameters a, $\delta$, and c are left free.  
From the fit, we obtained $a = 7.51 \pm 0.03$, $\delta = -0.033 \pm 0.005$, and $ c = 2.015 \pm 0.001$, in agreement with equation (\ref{eq:Moseley}). 
The results were then extrapolated to $Z > 37$. 

When the mean value and its error of the Gaussian modeling,
$32_{-3}^{+2}$~keV (table \ref{Fit Parameters}),
are considered to represent K$_\alpha$ emission line energies,
the responsible elements are identified
in  figure \ref{Energy vs. atomic num.} as $Z=59-63$,  or  Pr to Eu,
regardless of the assumed ionization states.
According to the calculation by Schatz et al (2001),
the rp-process has an endpoint at $Z=52$ (Te),
where further nucleosynthesis is prohibited by $\alpha$-decay process.
Then, the suggested elements would not be interpreted as rp-process products,
making the Gaussian modeling somewhat unphysical.
However, recent nuclear experiments  by Xing et al. (2018)
successfully refined the masses of $^{84}_{40}$Zr and $^{84}_{41}$Nb,
and updated  the structure of the nuclei involved in the rp-process.
As a result, the synthesis could proceed beyond the $Z=52$ endpoint.
Thus, the reality of the Gaussian interpretation must await
future studies in theoretical and experimental nuclear physics.

\subsubsection{redge model}\label{redge model}
\begin{figure*}[htb]
\begin{center}
\includegraphics[width=15cm,bb=50 50 800 600,clip]{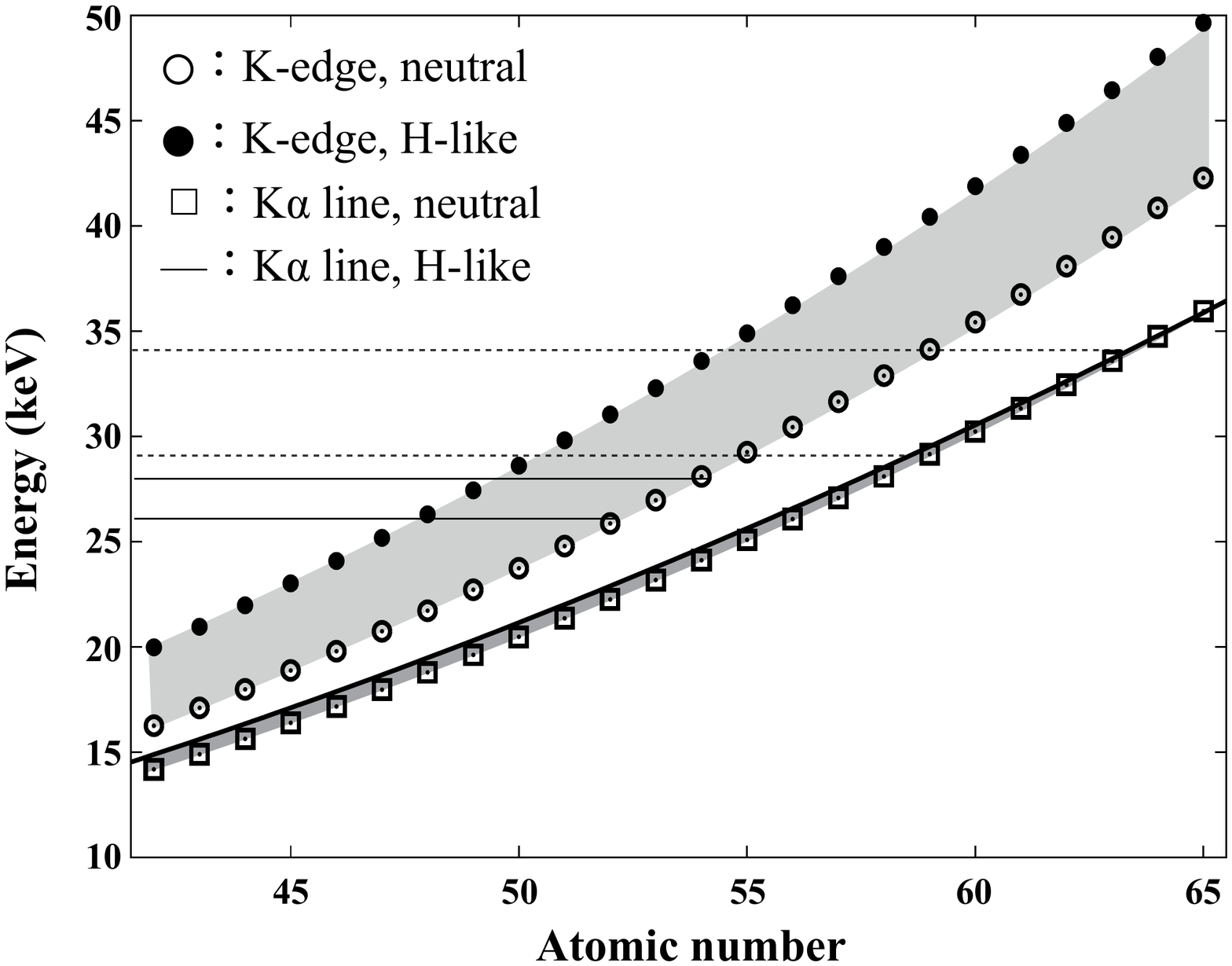}
\end{center}
\caption{
Energies of K$\alpha$ lines and K-edges of the elements with the atomic number from 40 to 65, 
assuming their production on the neutron star surface and different ionization states. 
The constraints from the Gaussian modeling is specified by a pair of horizontal dashed lines, 
and that from the redge modeling by a pair of solid lines.
}
\label{Energy vs. atomic num.}
\end{figure*}

In figure \ref{Energy vs. atomic num.},
the  filled circles  show the K-edge energies
of H-like ions as a function of $Z$,
taken from the NIST web page.
Although the figure also shows the case of neutral atoms
taken from the X-ray booklet \citep{XrayBooklet},
this is only for reference,
because K-shell ionization in a neutral atom would lead
predominantly to the K$_\alpha$ line emission
rather than recombination with a free electron. 

From the K-edge energy and its error ($27\pm1$~keV)
obtained with the redge modeling (table \ref{Fit Parameters}),
and assuming the H-like condition,
figure \ref{Energy vs. atomic num.} identifies
the corresponding elements as  $Z=48$ and 49 (Cd and In).
Since these are lower than the  $Z=52$ endpoint by  \citet{Schatz+2001},
the identification is allowed by the current understanding
of the rp-process nucleosynthesis in Type I bursts.
Considering the limited energy resolution of HXD-PIN, it is quite possible
that the feature is attributable to multiple elements near these atomic numbers,
rather than a single species.
In addition, the derived free-electron temperature of $11^{+10}_{-5}$ keV
(table \ref{Fit Parameters}) is reasonable as that in the atmosphere,
in the sense that it is in between those of the blackbody radiation
and the Comptonizing corona.

Although the redge interpretation is thus successful
up to this stage,
we are still left with the following questions
which are mainly of quantitative nature.
\begin{itemize}
\item[1] Whether Type I bursts can produce the responsible elements
plenty enough to explain the observed feature.
\item[2] How to sustain the synthesized heavy ions for a sufficiently long time
(e.g., days to weeks),  against their radioactive decays
and dilution by the accreting Hydrogen-rich materials.
\item[3] How to lift up the heavy elements up to
higher zones of the atmosphere,
where the Compton optical depth is sufficiently low as seen from us.

\item[4] Whether the heavy ions can be kept in the highly ionized condition
which is required by the redge interpretation.
\end{itemize}
These issues are however beyond the scope of the present paper.

\section{Conclusion} \label{Conclusion}
We analyzed  three broad-band Suzaku spectra of Aql X-1,
taken during the decay phase of an outburst in 2007 September to October.
The source was in the hard state,
with a 0.8--100.0 keV luminosity of $\sim 2 \times 10^{36}$ erg s$^{-1}.$
At about 30 keV of these spectra,
we detected a statistically significant excess-like feature
that cannot be explained by modifying the continuum models.
It may be considered as a K-shell feature of some heavy elements,
synthesized on the neutron-star surface
in thermonuclear flashes involving the rp-process.
More specifically,
the feature can be expressed as a K-emission line at about 32 keV,
or a recombination edge feature at an edge energy of 26 keV;
the former implies elements with $Z=59-63$ (Pr--Eu),
and the  latter $Z=48-49$ (Cd and In) assuming H-like ionization condition.
Although the former could be too high for the r-process products,
the latter is consistent with the currently accepted
r-process endpoints of $Z=52$.
To make this tentative interpretation more convincing,
we need quantitative evaluations from several aspects.


\begin{ack}
We would like to thank Dr. Liyi Gu for useful discussions. 
This work was partially supported by Grant-in-Aid for JSPS Fellows No. 16J05852.
\end{ack}

\appendix



\end{document}